\documentclass[%
 aip,
 amsmath,amssymb,
 reprint,%
]{revtex4-1}

\usepackage{graphicx}
\usepackage{dcolumn}
\usepackage{bm}
\usepackage{natbib}

\usepackage[utf8]{inputenc}
\usepackage[T1]{fontenc}
\usepackage{mathptmx}
\usepackage{etoolbox}
\usepackage{siunitx}

\DeclareSIUnit{\emu}{emu}
\makeatletter
\def\@email#1#2{%
 \endgroup
 \patchcmd{\titleblock@produce}
  {\frontmatter@RRAPformat}
  {\frontmatter@RRAPformat{\produce@RRAP{*#1\href{mailto:#2}{#2}}}\frontmatter@RRAPformat}
  {}{}
}%
\makeatother
\begin{document}

\preprint{AIP/123-QED}

\title{Magnetic signal scan imaging system based on giant magnetoimpedance (GMI) differential sensor}

\author{Tao Yang}
 \altaffiliation[Also at ]{Key Laboratory for Magnetism and Magnetic Materials of the Ministry of Education, Lanzhou University, Lanzhou, 730000, People's Republic of China.}
\author{Zhoulu Yu}
 \altaffiliation[Also at ]{Key Laboratory for Magnetism and Magnetic Materials of the Ministry of Education, Lanzhou University, Lanzhou, 730000, People's Republic of China.}
 \author{Xuekui Xi}
 \altaffiliation[Also at ]{Institute of Physics, Chinese Academy of Sciences, P.O. Box 603, Beijing 100190, People's Republic of China}
 \author{Changjun Jiang}
 \altaffiliation[Also at ]{Key Laboratory for Magnetism and Magnetic Materials of the Ministry of Education, Lanzhou University, Lanzhou, 730000, People's Republic of China.}
 \author{Guozhi Chai*}
 \email{chaigzh@lzu.edu.cn}
 \altaffiliation[Also at ]{Key Laboratory for Magnetism and Magnetic Materials of the Ministry of Education, Lanzhou University, Lanzhou, 730000, People's Republic of China.}

\date{\today} 

\begin{abstract}
This paper presents the design and implementation of a magnetic signal scanning and imaging system based on the giant magnetoimpedance (GMI) effect. The system employs a pair of performance-matched GMI sensing elements configured as a differential probe structure. Through co-optimized low-noise electronic and probe design, the system effectively suppresses both intrinsic sensor common-mode drift and external environmental magnetic noise, enabling high signal-to-noise ratio detection of nono-tesla to micro-tesla-level magnetic signals without magnetic shielding. Experimental results demonstrate that the differential system achieves significantly lower noise spectral density in unshielded environments compared to conventional GMI sensors (\SI{46}{pT}/$\sqrt{\text{Hz}}$ versus \SI{286}{pT}/$\sqrt{\text{Hz}}$ at \SI{1}{Hz}), with a sensitivity of 186,790 V/T and spatial resolution better than 200 micrometers. The system's excellent performance in weak magnetic field detection and spatial resolution was verified through scanning experiments of magnetic ink on US banknotes and magnetic reference samples. Compared to SQUID scanning systems, which requiring liquid helium cooling, this system based on the GMI effect offers advantages of room-temperature operation, compact structure, and low cost. Relative to conventional single-element GMI microscopes, it achieves significant improvements in signal-to-noise ratio and environmental adaptability. This research provides a practical solution for high-resolution magnetic field imaging at room temperature with broad application potential in materials magnetism, biomagnetic imaging, and nanomagnetic detection.
\end{abstract}

\maketitle

\section{\label{sec:level1}INTRODUCTION}

The precise measurement and microscopic imaging of magnetic fields are critically important in both fundamental scientific research and industrial applications. In particular, the growing demand for high-spatial-resolution scanning of weak magnetic fields has imposed increasingly stringent requirements on the sensitivity, resolution, and practicality of magnetic sensing technologies. Among existing magnetic sensors, the superconducting quantum interference device (SQUID) has long been regarded as the benchmark for weak magnetic field detection due to its ultrahigh sensitivity, which approaches fundamental physical limits\cite{10.1063/1.1448142,10.1063/1.1884025,GUDOSHNIKOV2020166938,10.1130/G21898.1}. SQUIDs operate based on the Josephson effect and magnetic flux quantization principles, but they must be maintained at liquid helium temperature (\SI{4.2}{K}) to function\cite{10.1063/1.5001390}. This cryogenic requirement leads to several inherent limitations. First, the complex cryogenic system necessitates a physical separation ranging from several micrometers to over one hundred micrometers between the SQUID sensor and room-temperature samples\cite{10.1063/1.1448142}\cite{10.1063/1.1884025}\cite{10.1063/1.1147570}\cite{https://doi.org/10.1029/2009GC002750}. Since magnetic field sensitivity decays with the cube or even the fourth power of distance, this separation severely limits the achievable spatial resolution, typically to the 10-\SI{100}{\um} range\cite{10.1063/1.1884025}\cite{10.1063/1.5001390}\cite{met13040800}\cite{https://doi.org/10.1029/2009GC002750}. Second, the stringent requirements for temperature stability, vibration isolation, and magnetic shielding result in high system complexity, substantial cost, and poor portability\cite{https://doi.org/10.1029/2009GC002750}\cite{10.1063/1.1142195}. 

Apart from SQUID, various magnetic sensing methodologies present distinct advantages and limitations. Tunnel magnetoresistance (TMR) sensors provide high sensitivity, compact dimensions, and room-temperature operation, yet exhibit higher background noise compared to SQUID systems\cite{annurev:/content/journals/10.1146/annurev.matsci.29.1.117}\cite{unknown}. Their sensitivity and noise characteristics in weak magnetic fields remain inadequate for measurements below the microtesla level, while also suffering from pronounced temperature drift and nonlinearity\cite{HONKURA2002375}. Nitrogen-vacancy (NV) center-based quantum magnetometers achieve atomic-scale spatial resolution and exceptional sensitivity, enabling nanoscale single-spin detection\cite{2020-49-6-002}. However, their effective operation requires complex laser initialization and microwave control systems, coupled with stringent demands on sample labeling, fluorescence collection efficiency, and NV center fabrication processes. These constraints limit their versatility and practicality in scanning imaging applications\cite{Zhao_2024}. Although conventional Hall sensors feature simple structure, good linearity, and easy integration, their relatively low sensitivity and significant noise generation generally restrict their application to magnetic field measurements above the millitesla level, making them unsuitable for direct weak-field imaging\cite{GUDOSHNIKOV2020166938}\cite{annurev:/content/journals/10.1146/annurev.matsci.29.1.117}\cite{hardware2040014,doi:10.1021/acsomega.2c02864,Collomb_2021}. Magnetic force microscopy (MFM) enables nanoscale spatial resolution by detecting magnetic interaction forces between probes and samples\cite{annurev:/content/journals/10.1146/annurev.matsci.29.1.117}\cite{PhysRevMaterials.3.074406}. However, as its operation relies on near-field force detection, measurements are susceptible to probe magnetization state, tip-sample distance, and non-magnetic interactions. This fundamentally limits its capability for absolute quantitative magnetic field measurement, with additional requirements for vacuum environments to mitigate noise\cite{https://doi.org/10.1029/2009GC002750}\cite{annurev:/content/journals/10.1146/annurev.matsci.29.1.117}\cite{Zhao_2024}.

Compared with the above-mentioned various types of magnetic sensors, magnetic sensors based on the giant magnetoimpedance (GMI) effect exhibit unique potential for practical applications\cite{GUDOSHNIKOV2020166938}\cite{HONKURA2002375}\cite{10.1063/1.112104,10.1063/1.358310,UCHIYAMA2020167148,s20010161}. The GMI effect refers to the significant change in the AC impedance of certain soft magnetic materials under high-frequency current excitation when exposed to an external magnetic field\cite{10.1063/1.112104}\cite{10.1063/1.358310}. GMI sensors based on this phenomenon offer advantages such as high linearity, fast response, room-temperature operation, and low cost\cite{HONKURA2002375}\cite{s20010161,MOHRI200185,10.1063/1.2722402}. However, conventional single-element GMI magnetic scanning microscopes remain limited by relatively high noise levels and weak resistance to environmental interference when measuring extremely weak magnetic fields\cite{GUDOSHNIKOV2020166938}\cite{s20010161}\cite{10.1063/1.2722402}\cite{8412532}. These systems usually require operation in a magnetically shielded environment, which restricts their broader adoption\cite{10.1063/1.1142195}.

To address the aforementioned technical limitations, this study developed a GMI-based magnetic scanning imaging system utilizing a differential architecture. The core innovation consists of a pair of performance-matched GMI sensing elements operating in a differential configuration. This design effectively suppresses both intrinsic sensor common-mode drift and external common-mode magnetic interference, enabling high signal-to-noise-ratio measurement of micro-tesla-level surface stray fields without magnetic shielding\cite{hardware2040014}\cite{UCHIYAMA2020167148}\cite{Takiya2017DevelopmentOA}. Experimental results demonstrate that the system achieves a lower noise floor and enhanced magnetic field resolution compared to conventional GMI microscopes. 

\begin{figure*}
    \centering
    \includegraphics[width=17cm]{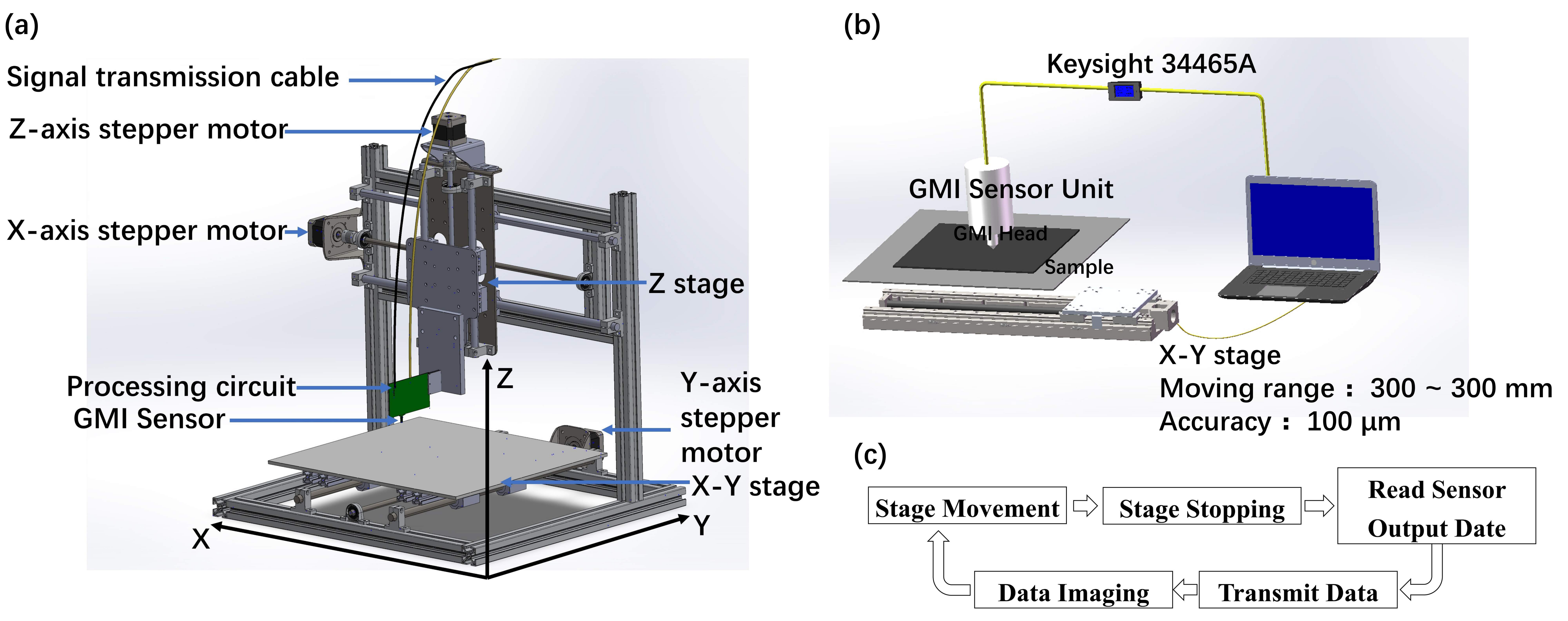}
    \caption{(a) Schematic diagram of the main components of the GMI-based differential magnetic signal scanning imaging system. The GMI differential magnetic sensor is mounted on a Z-axis displacement platform, while the sample is placed on an X-Y platform. The system operates without cooling or magnetic shielding. The three-axis displacement stage measures 46.0 cm in width, 39.0 cm in length, and 35.4 cm in height. (b) Configuration of the GMI differential magnetic sensor unit, showing the sensing probe mounted on a Z-axis displacement platform. (c) Schematic of the computer-controlled operation for the GMI differential magnetic scanning system. During magnetic field acquisition, the displacement stage remains stationary to minimize vibration-induced noise from the stepper motor.}
    \label{fig1}
\end{figure*}

\section{DESCRIPTION OF THE SCANNING GMI DIFFERENTIAL MAGNETIC SIGNAL IMAGING SYSTEM}

Figure \ref{fig1} presents a schematic diagram of the GMI differential magnetic signal scanning imaging system. Designed specifically for measuring weak magnetic fields from magnetic samples, the system employs non-magnetic materials such as brass, titanium and plastic in its construction to minimize magnetic interference and prevent influence on both the scanning equipment and the samples under test. The GMI differential magnetic sensor unit, comprising a differential magnetic sensor probe and a signal processing circuit board, is mounted on a Z-axis motorized translation stage. A plastic sample stage is installed on an aluminum platform, with the X-Y translation stages driven by stepper motors under host computer control.

Figure \ref{fig1}(b) illustrates the configuration of the GMI differential magnetic signal scanning imaging system. The GMI differential magnetic sensor unit, comprising a differential sensor probe and a matching circuit board, is mounted on a program-controlled Z-axis motorized stage with micrometer precision. With the controlling by the computer, the sensor unit can be positioned in close proximity to the magnetic sample surface, maintaining a probe-to-sample distance of less than \SI{200}{\um}. The plastic sample stage is assembled on a pure titanium lead screw, which is driven by computer-controlled X-Y stepper motors to enable sample positioning. An X-Y coordinate system marked on the stage surface facilitates programmed positioning prior to magnetic scanning. After positioning the differential magnetic probe at the target area, scanning is initiated. This design enables flexible scanning range configuration and precise sample repositioning for repeated experimental verification, which is essential for ensuring measurement reliability and eliminating random errors. The implemented differential magnetic scanning system provides scanning ranges of 300 mm and 200 mm along the X-axes and Y-axes respectively, with a resolution of \SI{10}{\um} and an accuracy of \SI{100}{\um}.

Figure \ref{fig1}(c) illustrates the operational principle of the GMI differential magnetic signal scanning imaging system. Sample movement is controlled by the host computer, and the translation stage halts after each step to eliminate magnetic interference generated during stage operation. The voltage signal from the sensor is acquired by a six-and-a-half-digit digital multimeter (Keysight 34465A), with digitized data transferred to the computer via a serial communication interface for subsequent magnetic field imaging. The scanning speed is set to 5 pixels per second. The step size of the translation stage corresponds to a magnetic pixel dimension ranging from \SI{10}{\um} to \SI{1000}{\um}. A typical 40 mm × 50 mm magnetic image with 200 × 250 pixels and a step size of \SI{200}{\um} can be acquired within approximately 3 hours. Both the X-Y-Z translation stages and the digital multimeter are automatically controlled by a program developed in LabVIEW.

\begin{figure*}
    \centering
    \includegraphics[width=17cm]{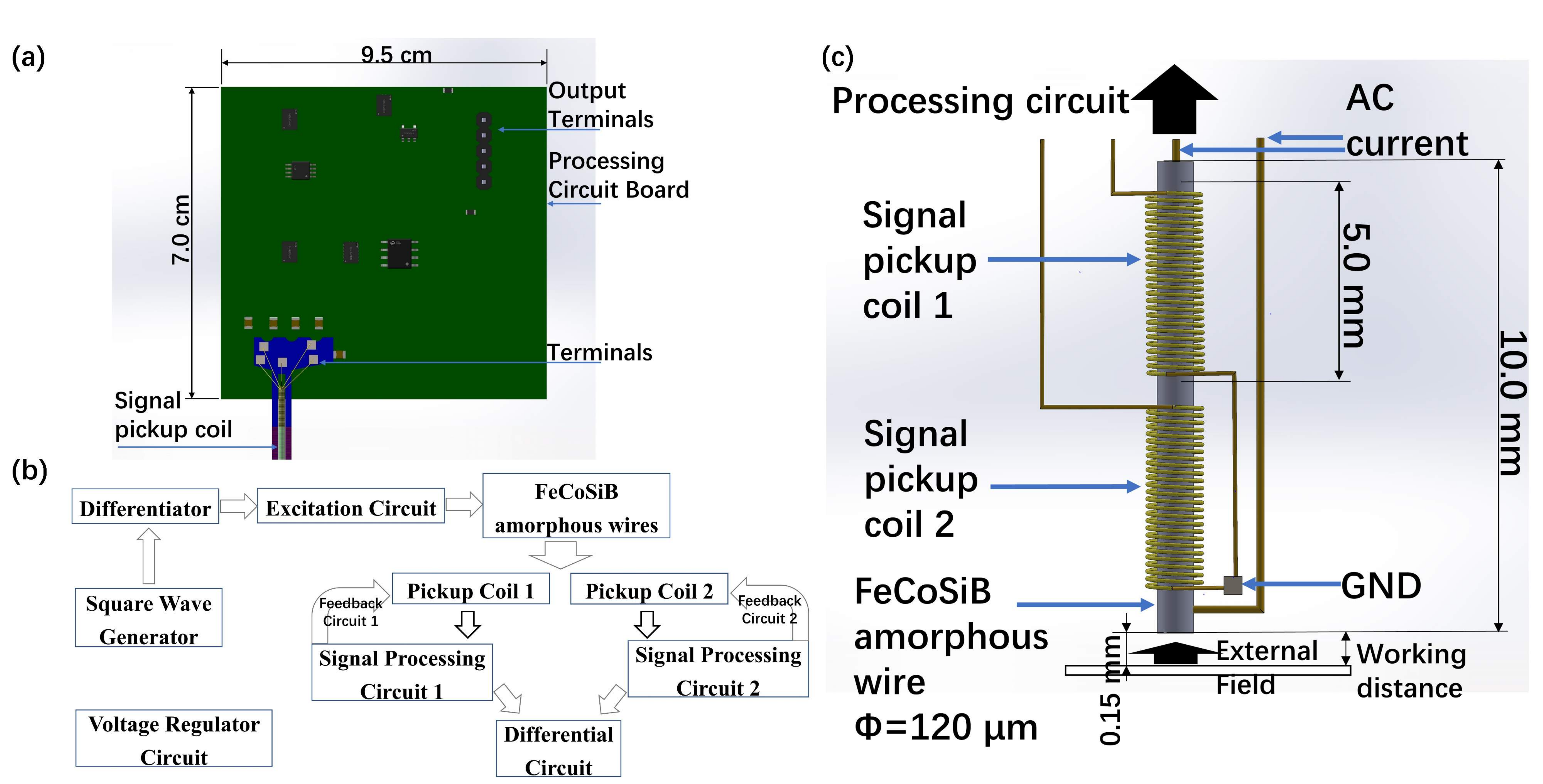}
    \caption{(a) Schematic of the three-dimensional model of the GMI differential magnetic sensing probe, mounted on a signal processing PCB. The PCB measures 9.5 cm in length and 7.0 cm in width, and the magnetic sensing probe is 1.5 cm in length. (b) Schematic of the signal processing circuit for the GMI differential magnetic sensor used in the scanning imaging system. (c) Structural diagram of the GMI differential magnetic sensing probe. The probe comprises an FeCoSiB amorphous wire with a diameter of \SI{120}{\um} and a length of 10 mm, along with two sets of coils, each 5.0 mm in length. The two coil sets are used to detect magnetic field variations on the surface of the amorphous wire and to provide both additive and negative feedback signals.}
    \label{fig2}
\end{figure*}

Meanwhile, the supporting circuitry and probe structure of the GMI differential magnetic sensor will be described in detail. Figure \ref{fig2}(a) illustrates a 3D schematic model of the GMI differential magnetic signal scanning sensor, which consists of a signal conditioning circuit board and a differential magnetic sensing probe. The circuit board measures 9.5 cm in length and 7.0 cm in width, and is equipped with both power supply and signal output ports. The GMI differential magnetic sensing probe has a length of 1.5 cm and incorporates six connection terminals. Among these, two terminals are used to supply excitation signals to the amorphous wire, while the remaining four are connected to two sets of sense coils, which detect the surface magnetic field of the amorphous wire and transmit the signals to the circuit board for processing. The signal conditioning circuit board and the sensing probe are interconnected via soldered joints.

Figure \ref{fig2}(b) presents a schematic diagram of the signal processing circuit for the GMI differential magnetic sensor used in the scanning imaging system. The circuit is primarily composed of two main sections: the excitation circuit and the signal processing circuit. The excitation circuit includes a square wave generator, a differentiator circuit, and a driving stage. The square wave generator produces a 100 kHz square wave with a 50\% duty cycle and a 5 V amplitude. This waveform is then shaped by the differentiator circuit to adjust its duty cycle. The final excitation stage employs an analog switch to enhance the current drive capability, thereby effectively exciting the FeCoSiB amorphous wire. The signal processing circuit comprises two identical signal channels, each containing a half-wave rectifier, a low-pass filter, and an integrator, along with a shared negative feedback loop and a differential output stage. Each channel processes the signal picked up by one of the sensing coils: the signal is first half-wave rectified, then filtered and integrated. The integrated output is fed back negatively to the corresponding sensing coil. After this processing, the signals from the two sensors are converted into two DC voltage levels. These DC signals are then differentially combined in the final stage to produce a single DC output voltage. This output voltage is divided by the magnetic-voltage conversion coefficient of the differential sensor (705770 V/T) to obtain the differential magnetic signal value at the measured sample point\cite{Takiya2017DevelopmentOA}.

Figure \ref{fig2}(c) illustrates the core component of the GMI differential magnetic sensor—the sensing probe. The probe consists of a custom-made FeCoSiB amorphous wire (\SI{120}{\um} in diameter, 10 mm in length) and two sets of sensing coils (each 5.0 mm long). Each coil set measures the magnetic field at its respective position. Through the accompanying signal processing circuit, the differential magnetic signal value is derived from the measurements of the two coil sets\cite{8412532}. 

\section{PERFORMANCE}

We characterized the performance of the GMI differential magnetic sensor through comparative analysis with a conventional GMI sensor structure, evaluating key parameters including sensitivity, noise level, and test methodology. Additionally, the developed GMI differential magnetic sensor demonstrated superior performance in scanning imaging experiments compared to traditional GMI magnetic sensors, exhibiting improvements in spatial resolution and weak magnetic field detection capability, as validated using standard samples such as US banknotes.

\begin{figure*}
    \centering
    \includegraphics[width=17cm]{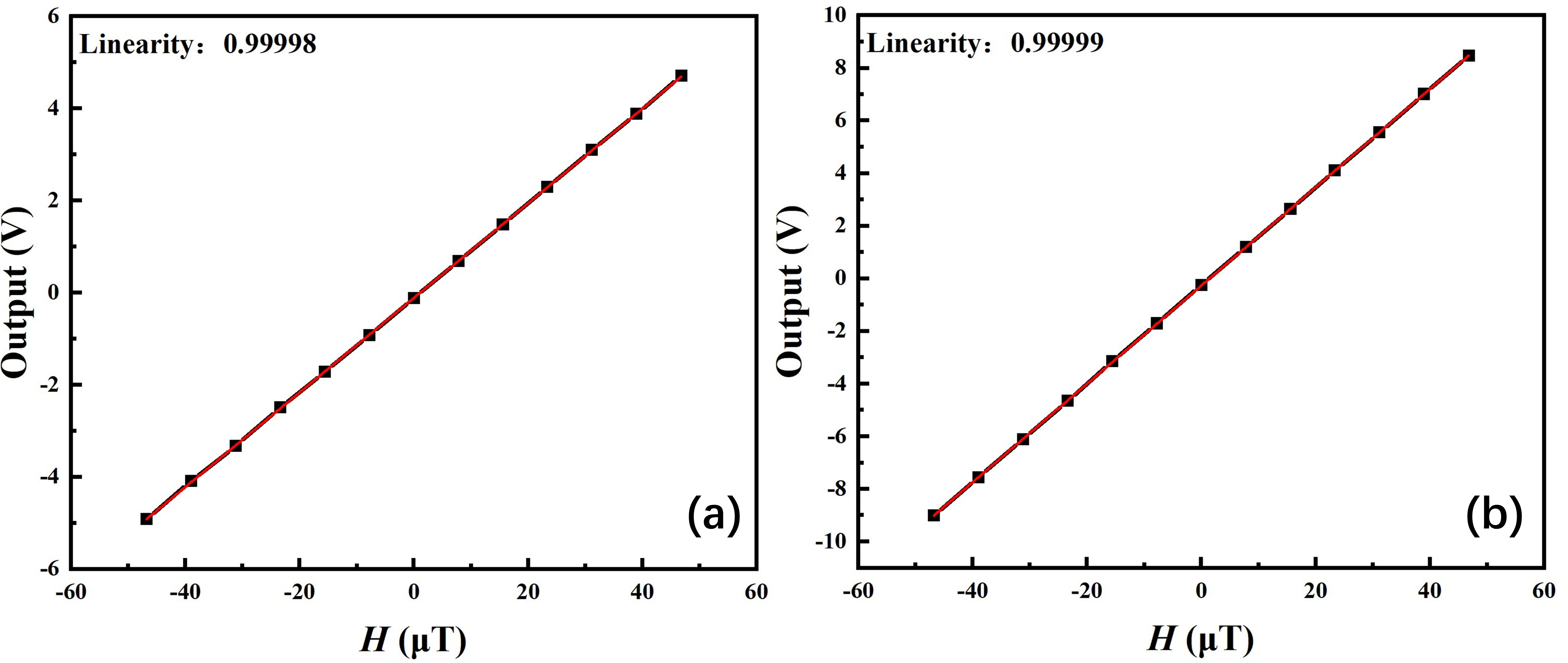}
    \caption{(a) Sensitivity characterization of the conventional GMI sensor. Under an external magnetic field sweeping from \SI{-46.8}{\micro T} to \SI{+46.8}{\uT}, the sensor output ranges from –4.917\,V to +4.706\,V. (b) Sensitivity characterization of the GMI differential magnetic sensor. Under the same magnetic field sweep from \SI{-46.8}{\micro T} to \SI{+46.8}{\uT}, the sensor output varies between –9.00\,V and +8.48\,V.}
    \label{fig3}
\end{figure*}

To quantitatively compare the performance of the GMI differential magnetic sensor and sensor types. The sensitivity of the conventional GMI sensor was measured as 102,650 V/T, as shown in Fig.\ref{fig3}(a). For the GMI differential magnetic sensor, the sensitivity of each channel was individually adjusted by tuning the feedback resistance to balance the responses of both channels. The sensitivity, as presented in Fig.\ref{fig3}(b), reached 186,790 V/T.

\begin{figure*}
    \centering
    \includegraphics[width=17cm]{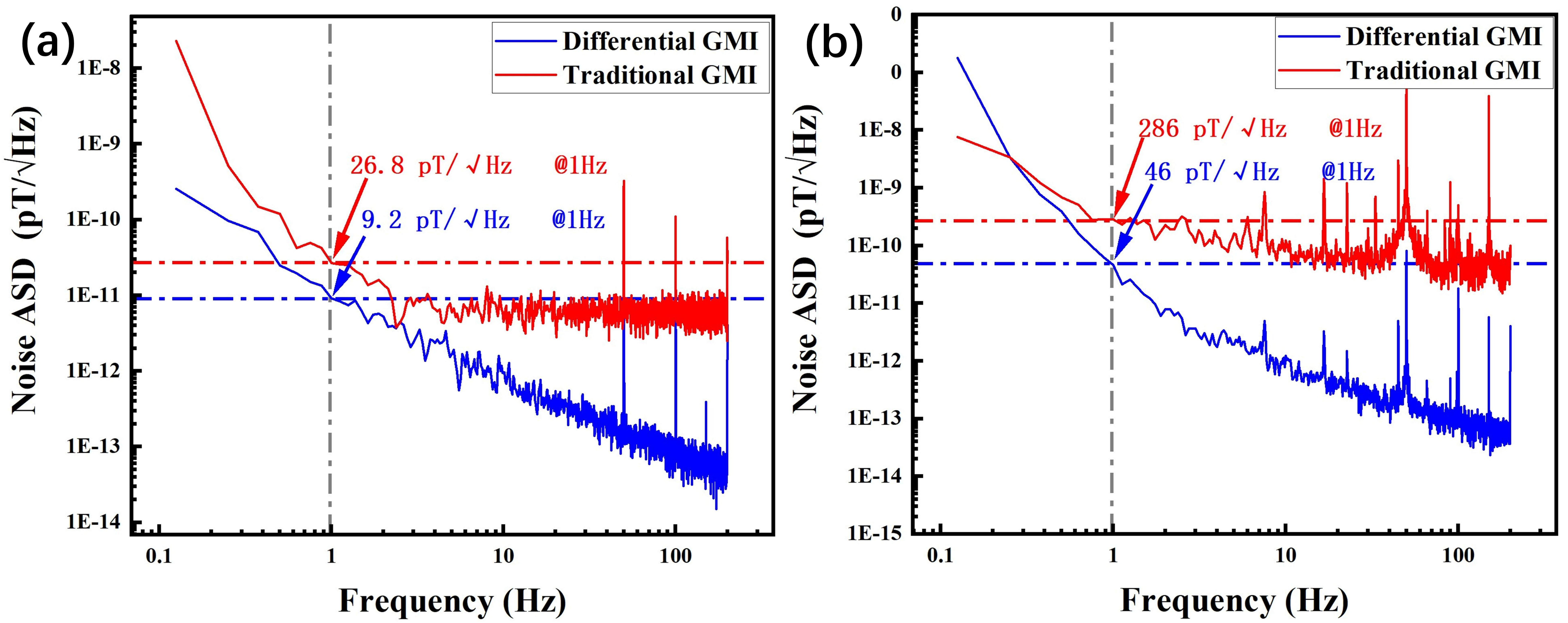}
    \caption{(a) Magnetic noise spectra of the GMI differential magnetic sensor and the conventional GMI magnetic sensor measured in a magnetically shielded environment. The measured noise spectral densities at 1 Hz are \SI{9.2}{pT}/$\sqrt{\text{Hz}}$ for the differential sensor and \SI{26.8}{pT}/$\sqrt{\text{Hz}}$ for the conventional sensor. (b) Magnetic noise spectra of both sensors acquired under unshielded conditions. The corresponding noise spectral densities at 1 Hz are \SI{46}{pT}/$\sqrt{\text{Hz}}$ (differential sensor) and \SI{286}{pT}/$\sqrt{\text{Hz}}$ (conventional sensor).}
    \label{fig4}
\end{figure*}

The magnetic noise performance of both sensors was evaluated under both magnetically shielded and unshielded conditions, as shown in Figs.\ref{fig4}(a) and \ref{fig4}(b). Figure \ref{fig4}(a) presents the magnetic noise test results for the two sensors in a shielded environment. The magnetic noise spectral densities at \SI{1}{Hz} are \SI{9.2}{pT}/$\sqrt{\text{Hz}}$ for the GMI differential magnetic sensor and \SI{26.8}{pT}/$\sqrt{\text{Hz}}$ for the conventional GMI magnetic sensor. These values are of the same order of magnitude, indicating comparable magnetic noise performance between the two sensors under magnetically shielded conditions. Figure \ref{fig4}(b) displays the magnetic noise spectra of the two sensors measured in an unshielded environment. Here, the noise spectral densities at \SI{1}{Hz} are \SI{46}{pT}/$\sqrt{\text{Hz}}$ for the GMI differential sensor and \SI{286}{pT}/$\sqrt{\text{Hz}}$ for the conventional sensor. These results clearly demonstrate that the GMI differential magnetic sensor exhibits significantly lower magnetic noise than the conventional sensor in unshielded conditions. This outcome highlights the superior immunity of the differential sensor to environmental magnetic interference and suggests that it will perform more effectively in practical scanning applications.

Description of the testing method and data processing: The sensitivity was characterized by placing the sensor within the uniform magnetic field region of a pre-calibrated solenoid inside a magnetic shielding enclosure. By applying varying currents to the solenoid, known magnetic fields were generated. The corresponding output voltage of the sensor was recorded for each applied field, and the sensitivity was subsequently calculated from the recorded voltage–magnetic field relationship.

Magnetic noise characterization procedure: Magnetic noise testing was performed using high-precision, low-noise data acquisition systems to record sensor output inside a magnetically shielded environment. The acquired time-domain data were then processed via Fast Fourier Transform (FFT) to obtain the voltage noise power spectral density. The magnetic noise spectral density was subsequently derived by taking the square root of the power spectral density and normalizing it by the sensor's sensitivity.

\begin{figure}
    \centering
    \includegraphics[width=8.5cm]{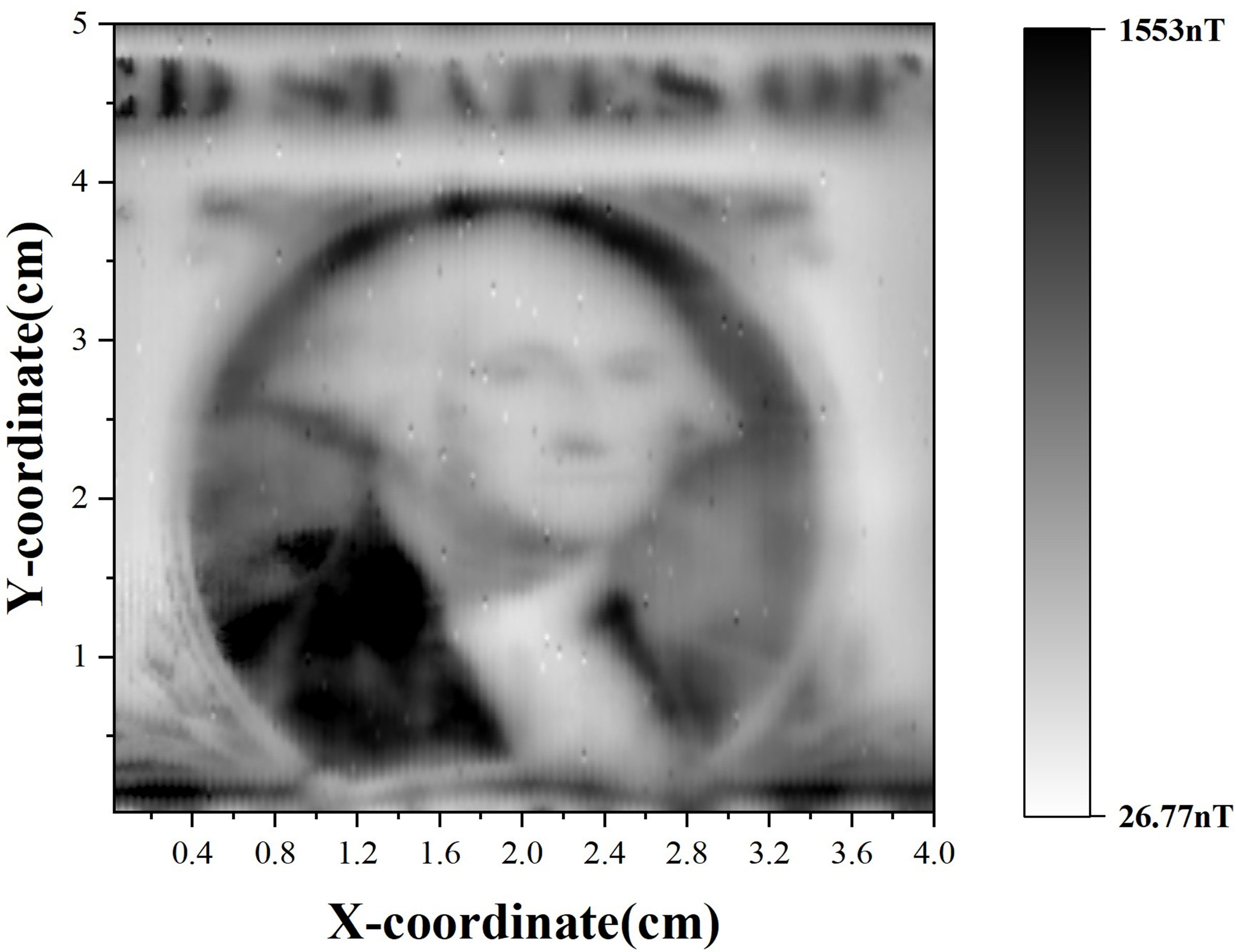}
    \caption{Magnetic field image of a United States one-dollar note section acquired at a working distance of \SI{150}{\um}. The banknote was previously magnetized by a 4.0 T vertically upward magnetic field. The grayscale represents magnetic field intensities ranging from approximately 26.77 nT (downward-pointing, white) to 1553 nT (upward-pointing, black).}
    \label{fig5}
\end{figure}

The spatial resolution of the GMI differential magnetic scanning imaging system was effectively evaluated by scanning the magnetic ink pattern of George Washington's portrait on a one-dollar bill\cite{10.1063/1.1147570}\cite{10.1063/1.2722402}. The GMI differential sensor was positioned at a working distance of approximately \SI{150}{\um} from the banknote surface. The scanning area was configured with X- and Y-axis dimensions of 40 mm and 50 mm respectively, with scan step sizes set to 0.2 mm in both directions. The resulting magnetic image, presented in Fig.\ref{fig5}, clearly reveals facial features and structural details of the portrait. Specifically, the chin outline, collar elements, and the floral motifs on both sides of the head are well resolved. Comparative analysis with optical images indicates that the spatial resolution of the magnetic imaging system exceeds 200 micrometers.

\begin{figure*}
    \centering
    \includegraphics[width=17cm]{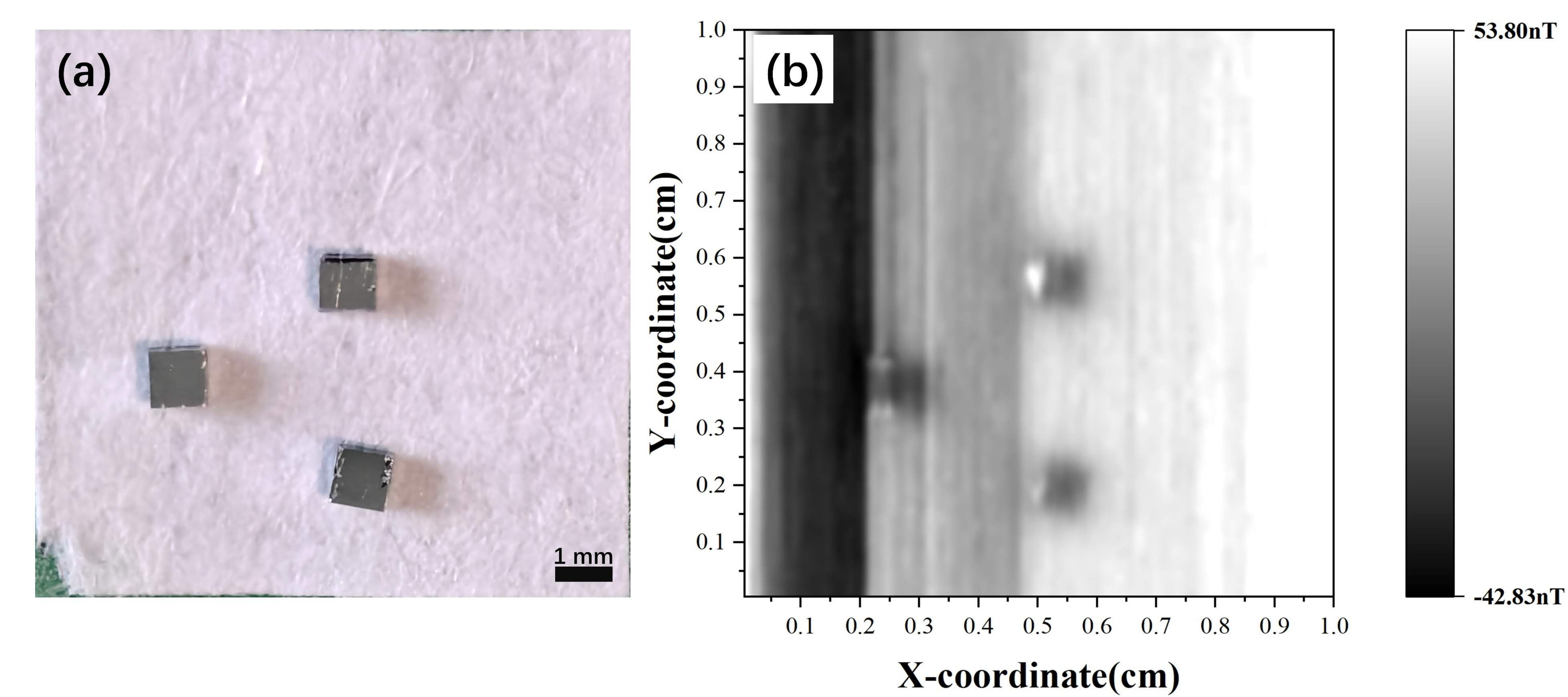}
    \caption{(a) Physical layout of three magnetic reference samples, each measuring 1 mm × 1 mm with a magnetic moment of \SI{3}{\micro\emu}, distributed on a \SI{1}{cm}×\SI{1}{cm} plane. (b) Magnetic field image acquired with the differential magnetic sensing probe positioned approximately \SI{150}{\um} above the sample plane. The image captures the combined magnetic response under the gradient field generated by the X-axis stepper motor. The grayscale represents magnetic field values ranging from \SI{-42.83}{nT} (upward-pointing, black) to \SI{53.8}{n T} (downward-pointing, white).).}
    \label{fig6}
\end{figure*}

To validate the capability of the GMI differential magnetic scanning imaging system in detecting weak magnetic signals and establishing a one-to-one correspondence between the distribution of magnetic reference samples and their surface magnetic field patterns, We fabricated three identical magnetic reference samples, each measuring \SI{1}{mm} $\times$ \SI{1}{mm}. Each sample surface was deposited with a multilayer structure comprising four alternating layers of \SI{0.5}{\um} cobalt (Co) and \SI{0.8}{\um} platinum (Pt), with a measured magnetic moment of \SI{3}{\micro\emu} per sample. These samples were randomly arranged within a \SI{1}{cm} $\times$ \SI{1}{cm} planar area and scanned at a working distance of approximately \SI{150}{\um}. The experimental setup presents additional complexity due to the significant non-uniform gradient magnetic field generated by the X-Y axis stepper motors. During operation, while the GMI differential magnetic sensor moves with the Z-axis stepper motor, the X-Y axis stepper motors remain stationary. However, the X-axis stepper motor drives the sensor assembly (including the Z-axis mechanism) via a lead screw, causing the relative position between the magnetic sensor and the X-Y axis motors to continuously change. Consequently, the sensor measurements incorporate the gradient field produced by these motors. This configuration creates a challenging test scenario aimed at studying magnetic images of three identical reference samples within the non-uniform gradient magnetic field environment generated by the X-Y axis stepper motors.This results in a situation where the scanned images do not directly represent the distribution of the magnetic reference samples; rather, they provide a magnetic field map that clearly displays the superimposed magnetic field resulting from the combination of the samples' magnetic fields and the non-uniform gradient magnetic field. Figure \ref{fig6} shows the magnetic field image across the \SI{1}{cm} $\times$ \SI{1}{cm} scanning area containing the reference samples under this gradient field. In a uniform magnetic environment, the stray field distribution of a reference sample directly reflects its intrinsic magnetic properties. However, in the presence of a background gradient field, the signal detected by the microscope probe results from the vector superposition of the magnetic field generated by the sample and the spatially varying background gradient. The contrast of each pixel in the observed image is thus jointly influenced by both contributions. Although the three reference samples are intrinsically identical, they exhibit clearly distinguishable features in the final scanned image due to their different positions within the background gradient field. Samples located in regions with a stronger background field may experience partial polarization or pre-orientation of their magnetic moments, leading to shifts in domain walls or alterations in their magnetization state. When the sample's stray field superimposes with a stronger background field, significant signal enhancement or attenuation may occur (depending on their relative orientation), manifesting as abnormally bright or dark contrast in the image. In contrast, samples situated in weaker background field regions exhibit signals closer to their intrinsic magnetic domain structure, though still affected by an overall shift in grayscale due to the background gradient. These observations demonstrate that our GMI differential magnetic scanning imaging system can detect magnetic reference samples featuring a \SI{1}{mm} $\times$ \SI{1}{mm} planar structure with four alternating layers of \SI{0.5}{\um} cobalt and \SI{0.8}{\um} platinum, each possessing a magnetic moment of \SI{3}{\micro\emu}, even under gradient magnetic field conditions.

\section{DISCUSSION}
The GMI differential magnetic signal scanning imaging system developed in this work demonstrates substantial innovation in both structural design and performance parameters. By employing a differential sensing configuration, the system effectively suppresses environmental interference and intrinsic sensor common-mode drift, enabling high signal-to-noise detection of microtesla-level magnetic signals without magnetic shielding. This capability permits spatial distribution imaging of weak magnetic fields in standard laboratory environments, significantly expanding the application scope of GMI sensing technology.

However, the spatial resolution of the current system remains constrained by the sensor-to-sample spacing and the physical dimensions of the amorphous wire. The thickness of the existing protective coating around the amorphous wire (approximately \SI{200}{\um}), along with variations in lift-off distance caused by surface topography of samples, influences the accuracy of magnetic signal measurement. Future improvements may focus on the following aspects: (1) Microstructural optimization of the sensor: employing micro-nano fabrication techniques to shape FeCoSiB amorphous wires into tapered tip structures, thereby reducing the effective sensing area and enhancing local magnetic response sensitivity; (2) Real-time lift-off distance feedback: integrating optical or capacitive distance measurement modules to enable dynamic calibration of the sensor-sample separation during scanning; (3) Enhanced signal processing algorithms: incorporating digital filtering and multi-dimensional differential algorithms to further suppress systematic drift errors; (4) Multi-probe array deployment: extending the differential probe configuration into a two-dimensional array architecture to improve scanning throughput and spatial coverage.

From a broader perspective, this system establishes a new balance between performance and practicality. Compared with SQUID microscopy that requires liquid helium cooling and high-grade magnetic shielding, the GMI differential magnetic scanning imager demonstrates significant advantages including room-temperature operation, miniaturized design, and high stability. Meanwhile, its weak-field detection capability and spatial resolution are approaching that of SQUID systems. Consequently, this technology not only shows substantial value for fundamental physics and magnetic materials research, but also provides practical engineering solutions for applications such as geological surveying, biomagnetic imaging, and magnetic micro-device inspection.

\section{CONCLUSION}
This study presents the design and implementation of a magnetic signal scanning and imaging system based on the differential giant magnetoimpedance (GMI) effect. Through collaborative optimization of the differential probe structure and low-noise electronic design, the system achieves high-sensitivity detection and high-resolution imaging of micro-tesla-level magnetic signals without magnetic shielding. Experimental results demonstrate the system's significant advantages in spatial resolution, interference immunity, and imaging accuracy. Compared with conventional unitary GMI scanners, the developed system shows marked improvement in signal-to-noise ratio and environmental adaptability. Relative to SQUID-based systems, it substantially reduces both cost and operational complexity. This research offers a viable alternative for high-resolution magnetic field imaging at room temperature and establishes a solid foundation for future applications in biomagnetism, materials science, and nanomagnetic detection. In summary, the developed differential GMI microscope effectively integrates the practical benefits of GMI technology with the performance advantages of differential sensing\cite{10.1063/1.112104}\cite{10.1063/1.358310}\cite{MOHRI200185}. While maintaining the compact dimensions and room-temperature operation of conventional GMI sensors, it substantially extends performance boundaries, delivering a solution that combines high sensitivity, strong interference rejection, and excellent engineering practicality for high-resolution weak magnetic field scanning and imaging\cite{hardware2040014}\cite{10416987}.

\begin{acknowledgments}
This work was supported in part by the National Key Research and Development Program of China under Grant 2024YFF0726700 and the National Natural Science Foundation of China (NSFC) (Nos. 52471200, 12174165 and 52201219).

\end{acknowledgments}

\bibliography{aipsamp}

\begin{thebibliography}{10}

\bibitem{10.1063/1.1448142}
F.~Baudenbacher, N.~T. Peters, and Jr. Wikswo, J.~P.
\newblock High resolution low-temperature superconductivity superconducting quantum interference device microscope for imaging magnetic fields of samples at room temperatures.
\newblock {\em Review of Scientific Instruments}, 73(3):1247--1254, 03 2002.

\bibitem{10.1063/1.1884025}
L.~E. Fong, J.~R. Holzer, K.~K. McBride, E.~A. Lima, F.~Baudenbacher, and M.~Radparvar.
\newblock High-resolution room-temperature sample scanning superconducting quantum interference device microscope configurable for geological and biomagnetic applications.
\newblock {\em Review of Scientific Instruments}, 76(5):053703, 04 2005.

\bibitem{GUDOSHNIKOV2020166938}
S.~Gudoshnikov, V.~Tarasov, B.~Liubimov, V.~Odintsov, S.~Venediktov, and A.~Nozdrin.
\newblock Scanning magnetic microscope based on magnetoimpedance sensor for measuring of local magnetic fields.
\newblock {\em Journal of Magnetism and Magnetic Materials}, 510:166938, 2020.

\bibitem{10.1130/G21898.1}
J\'er\^ome Gattacceca, Michel Boustie, Benjamin~P. Weiss, Pierre Rochette, Eduardo~A. Lima, Luis~E. Fong, and Franz~J. Baudenbacher.
\newblock Investigating impact demagnetization through laser impacts and squid microscopy.
\newblock {\em Geology}, 34(5):333--336, 05 2006.

\bibitem{10.1063/1.5001390}
P.~Reith, X.~Renshaw~Wang, and H.~Hilgenkamp.
\newblock Analysing magnetism using scanning squid microscopy.
\newblock {\em Review of Scientific Instruments}, 88(12):123706, 12 2017.

\bibitem{10.1063/1.1147570}
Thomas~S. Lee, Eugene Dantsker, and John Clarke.
\newblock High‐transition temperature superconducting quantum interference device microscope.
\newblock {\em Review of Scientific Instruments}, 67(12):4208--4215, 12 1996.

\bibitem{https://doi.org/10.1029/2009GC002750}
Fatim Hankard, Jérôme Gattacceca, Claude Fermon, Myriam Pannetier-Lecoeur, Benoit Langlais, Yoann Quesnel, Pierre Rochette, and Suzanne~A. McEnroe.
\newblock Magnetic field microscopy of rock samples using a giant magnetoresistance–based scanning magnetometer.
\newblock {\em Geochemistry, Geophysics, Geosystems}, 10(10), 2009.

\bibitem{met13040800}
Georgy Danilov, Yury Grebenshchikov, Vladimir Odintsov, Margarita Churyukanova, and Sergey Gudoshnikov.
\newblock Measurements of stray magnetic fields of fe-rich amorphous microwires using a scanning gmi magnetometer.
\newblock {\em Metals}, 13(4), 2023.

\bibitem{10.1063/1.1142195}
Yu~Pei Ma and Jr. Wikswo, John~P.
\newblock Magnetic shield for wide‐bandwidth magnetic measurements for nondestructive testing and biomagnetism.
\newblock {\em Review of Scientific Instruments}, 62(11):2654--2661, 11 1991.

\bibitem{annurev:/content/journals/10.1146/annurev.matsci.29.1.117}
John~R. Kirtley and John~P. Wikswo.
\newblock Scanning squid microscopy.
\newblock {\em Annual Review of Materials Research}, 29(Volume 29, 1999):117--148, 1999.

\bibitem{unknown}
Hirokuni Oda, Seiji Kumagai, Kosuke Fujiwara, Hitoshi Matsuzaki, Hiroshi Wagatsuma, Mikihiko Oogane, Hitoshi Kubota, Naoto Fukuyo, and Akihiro Tanimoto.
\newblock Advancement in scanning magnetic microscopy utilizing high-sensitivity room-temperature tmr sensors for geological applications, 08 2024.

\bibitem{HONKURA2002375}
Yoshinobu Honkura.
\newblock Development of amorphous wire type mi sensors for automobile use.
\newblock {\em Journal of Magnetism and Magnetic Materials}, 249(1):375--381, 2002.
\newblock International Workshop on Magnetic Wires.

\bibitem{2020-49-6-002}
DING Zhe, SHI Fa-Zhan, and DU~Jiang-Feng.
\newblock Nanoscale magnetic imaging based on quantum sensing with diamond and its applications to condensed matter physics.
\newblock {\em PHYSICS}, 49(6):359--372, 2020.

\bibitem{Zhao_2024}
Xu-Tong Zhao, Fei-Yue He, Ya-Wen Xue, Wen-Hao Ma, Xiao-Han Yin, Sheng-Kai Xia, Ming-Jing Zeng, and Guan-Xiang Du.
\newblock High-resolution imaging of magnetic fields of banknote anti-counterfeiting strip using fiber diamond probe.
\newblock {\em Chinese Physics B}, 33(4):048502, mar 2024.

\bibitem{hardware2040014}
Artem Sobko, Nikolai Yudanov, Larissa~V. Panina, and Valeriya Rodionova.
\newblock The development of a 3d magnetic field scanner using additive technologies.
\newblock {\em Hardware}, 2(4):279--291, 2024.

\bibitem{doi:10.1021/acsomega.2c02864}
Tongyu Dai, Hua Xu, Shanshan Chen, and Zhiyong Zhang.
\newblock High performance hall sensors built on chemical vapor deposition-grown bilayer graphene.
\newblock {\em ACS Omega}, 7(29):25644--25649, 2022.

\bibitem{Collomb_2021}
David Collomb, Penglei Li, and Simon Bending.
\newblock Frontiers of graphene-based hall-effect sensors.
\newblock {\em Journal of Physics: Condensed Matter}, 33(24):243002, may 2021.

\bibitem{PhysRevMaterials.3.074406}
Qiyuan Feng, Dechao Meng, Haibiao Zhou, Genhao Liang, Zhangzhang Cui, Haoliang Huang, Jianlin Wang, Jinghua Guo, Chao Ma, Xiaofang Zhai, Qingyou Lu, and Yalin Lu.
\newblock Direct imaging revealing halved ferromagnetism in tensile-strained $\mathrm{LaCo}{\mathrm{o}}_{3}$ thin films.
\newblock {\em Phys. Rev. Mater.}, 3:074406, Jul 2019.

\bibitem{10.1063/1.112104}
L.~V. Panina and K.~Mohri.
\newblock Magneto‐impedance effect in amorphous wires.
\newblock {\em Applied Physics Letters}, 65(9):1189--1191, 08 1994.

\bibitem{10.1063/1.358310}
L.~V. Panina, K.~Mohri, K.~Bushida, and M.~Noda.
\newblock Giant magneto‐impedance and magneto‐inductive effects in amorphous alloys (invited).
\newblock {\em Journal of Applied Physics}, 76(10):6198--6203, 11 1994.

\bibitem{UCHIYAMA2020167148}
Tsuyoshi Uchiyama and Jiaju Ma.
\newblock Development of pico tesla resolution amorphous wire magneto-impedance sensor for bio-magnetic field measurements.
\newblock {\em Journal of Magnetism and Magnetic Materials}, 514:167148, 2020.

\bibitem{s20010161}
Dongfeng He.
\newblock Pt-level high-sensitivity magnetic sensor with amorphous wire.
\newblock {\em Sensors}, 20(1), 2020.

\bibitem{MOHRI200185}
K.~Mohri, T.~Uchiyama, L.P. Shen, C.M. Cai, and L.V. Panina.
\newblock Sensitive micro magnetic sensor family utilizing magneto-impedance (mi) and stress-impedance (si) effects for intelligent measurements and controls.
\newblock {\em Sensors and Actuators A: Physical}, 91(1):85--90, 2001.
\newblock Third European Conference on Magnetic Sensors \& Actuators.

\bibitem{10.1063/1.2722402}
Minoru Uehara and Norihiro Nakamura.
\newblock Scanning magnetic microscope system utilizing a magneto-impedance sensor for a nondestructive diagnostic tool of geological samples.
\newblock {\em Review of Scientific Instruments}, 78(4):043708, 04 2007.

\bibitem{8412532}
Jiaju Ma and Tsuyoshi Uchiyama.
\newblock Development of peak-to-peak voltage detector-type mi gradiometer for magnetocardiography.
\newblock {\em IEEE Transactions on Magnetics}, 54(11):1--5, 2018.

\bibitem{Takiya2017DevelopmentOA}
Takashi Takiya and Tsuyoshi Uchiyama.
\newblock Development of active shielding-type mi gradiometer and application for magnetocardiography.
\newblock {\em IEEE Transactions on Magnetics}, 53:1--4, 2017.

\bibitem{10416987}
Ruixuan Yao and Tsuyoshi Uchiyama.
\newblock Analysis of magnetic signatures for vehicle detection using dual-axis magneto-impedance sensors.
\newblock {\em IEEE Sensors Journal}, 24(6):8721--8730, 2024.

\end{thebibliography}

\end{document}